\documentstyle[12pt]{article}

\begin{document}

\hoffset = -1truecm
\voffset = -2truecm
\title{\bf Effect of hydrogen on ground state structures 
of  small silicon clusters }
\author{
{D. Balamurugan }
\hspace{0.05cm}
{\normalsize and }
{ R. Prasad }
\\
\em 
{Department of Physics,} 
\\
{Indian Institute of Technology, Kanpur 208016, India}
}
\newpage
\maketitle
\begin{abstract}

We present results for  ground state structures   
of small
Si$_{n}$H (2\( \leq  \)\emph{n} \( \leq  \)10) clusters 
using the Car-Parrinello molecular dynamics. In particular, we 
focus on how  the addition  of a  hydrogen atom affects the
ground state geometry, total energy and the first excited 
electronic level  gap of an Si$_{n}$ cluster.  We discuss  the 
nature of bonding of hydrogen in these clusters.  We find that 
hydrogen  bonds with two silicon atoms only in Si$_{2}$H, Si$_{3}$H 
and Si$_{5}$H clusters, while in other clusters 
(\emph{i.e.} Si$_{4}$H, Si$_{6}$H, Si$_{7}$H, Si$_{8}$H,
Si$_{9}$H and Si$_{10}$H) hydrogen  is bonded to only 
one silicon atom. Also in the case of a  compact and closed 
silicon cluster hydrogen bonds to the cluster 
from outside. We find that the first excited 
electronic level gap  of Si$_{n}$ and Si$_{n}$H fluctuates 
as a function of size and  this may provide a first principles 
basis for the short-range potential fluctuations in 
hydrogenated amorphous silicon. Our results show that the 
addition of a single hydrogen can cause large changes in  
the electronic structure  of  a silicon cluster, though 
the geometry is not much affected. Our 
calculation of  the lowest energy  fragmentation 
products of Si$_{n}$H  clusters 
shows that hydrogen is easily removed from Si$_{n}$H clusters.
\end{abstract}

\vskip 1in
\newpage

\vskip .2in
\par
\indent

\section{INTRODUCTION}

During the last decade clusters have attracted a 
lot of  attention  because of their interesting  
 and novel properties\cite{kn:sm}. Of particular interest 
are the binary clusters of hydrogen and silicon which are 
thought to be present in hydrogenated amorphous 
silicon(\emph{a}-Si:H),  porous silicon and
silicon surfaces. In addition to  the fundamental
interest, their study may throw some light on 
complex phenomena occurring in these systems.  Hydrogen plays
an important role in these systems in phenomena like 
photoluminescence of porous silicon, potential fluctuations and 
the Staebler-Wronski effect in hydrogenated amorphous 
silicon(\emph{a}-Si:H)\cite{kn:pra1}-\cite{kn:sca3}. To 
understand these phenomena it is important to study how 
the addition of hydrogen affects the 
local electronic structure and geometry in these 
systems\cite{kn:pra1}, \cite{kn:fri}. Since these systems 
are very difficult to handle computationally, some understanding in
this regard can be gained by simpler calculations on small 
hydrogenated silicon  clusters. With this motivation, we have 
carried out 
a detailed study of ground state structures and 
electronic properties of small Si$_{n}$H
clusters( 2\protect\( \leq \emph{n}\leq 10\protect \)) using the 
Car-Parrinello molecular dynamics(CPMD), focusing particularly 
on  the effects caused by  hydrogen. In our study
we have investigated: (1) ground state geometries 
 of Si$_{n}$H clusters, (2) effect of hydrogen on the
geometry of a silicon cluster, (3)  stability of a silicon cluster 
due to the addition of hydrogen, (4) the first excited electronic 
level gaps of Si$_{n}$H and Si$_{n}$
clusters, (5) bonding nature and position
of hydrogen in silicon clusters and  (6) the lowest energy
fragmentation products of Si$_n$H  and Si$_{n}$ clusters. 

Several calculations have been done for many silicon 
hydrogen clusters by using various techniques.  Using 
the Car-Parrinello method, Onida and Andreoni\cite{kn:and} 
studied the ground state geometry and electronic structure 
of hydrogen passivated crystalline fragments of silicon 
such as Si$_{5}$H$_{12}$, Si$_{6}$H$_{16}$, Si$_{8}$H$_{8}$,
Si$_{10}$H$_{16}$ and Si$_{14}$H$_{24}$. They
found that Si-Si bond lengths were  insensitive to size effects, 
but electronic properties were  strongly affected.  They also
 found that HOMO-LUMO gaps were not simply related to the size 
of the clusters and  the localization of electronic states 
near the gaps was not necessarily silicon-like, even 
though the clusters are crystal fragments of silicon passivated by
hydrogen. Quantum chemical calculations of Si$_{n}$H$_{m}$
clusters were carried out by Meleshko \emph{et al}\cite{kn:mel} 
for \emph{n}=6-16 and \emph{m} ranging from 2 to 20. They 
found that each H atom was bonded with only one silicon
atom and localized outside the silicon skeleton and  that the 
packing density in the skeleton decreased as the hydrogen 
content of the cluster increased.  Miyazaki \emph{et al}\cite{kn:tak} 
performed density functional calculations for 
small hydrogenated silicon clusters
of Si$_{6}$H$_{x}$ (0\protect\( \leq \emph{x}\leq \protect \)14)  
and showed that for the sequence 
Si$_{6}$H$_{x-2}$ + H$_{2}$ $\rightarrow $ Si$_{6}$H$_{x}$, 
 the attachment of H occurred not at the site of silicon 
having dangling bonds but at the site where the LUMO of Si$_{6}$H$_{x}$
has a large amplitude for \emph{x}=2 and 6. According to 
this calculation, the bonding interaction of \emph{1s} orbitals
of hydrogen atoms  with the LUMO of Si$_{6}$H$_{x-2}$
should be the major cause of stabilization of the clusters. Their 
explanation makes it clear that hydrogen does not simply 
attach with silicon to saturate the dangling bonds, but it 
interacts at electronic level.

Swihart \emph{et al}\cite{kn:mar1} used  \emph{ab initio} molecular
orbital calculations to investigate structure and energetics
of selected hydrogenated silicon clusters containing six to 
ten silicon atoms.  The clusters investigated were those that
played the most important role in particle nucleation\cite{kn:mar2} 
in silane during chemical vapor  
deposition. Shvartsburg \emph{et al} \cite{kn:shv} 
modeled the dissociation of neutral and positively 
charged Si$_{n}$ clusters in  \emph{n}\protect\( \leq \protect \)26 
range. They used dissociation energies to test the results 
of global optimization and  fragmentation products of the
clusters.  Recently, non-orthogonal tight-binding molecular
dynamics(NTBMD) with simulated annealing optimization method was  
used to calculate  ground state geometries of small Si$_{n}$H
clusters\cite{kn:gup} (2\protect\( \leq \emph{n}\leq \protect \)10) 
and Si$_{2}$H$_{m}$(\emph{n}=1, 2 and \emph{m}=2-6) 
clusters\cite{kn:gup2}, \cite{kn:foot}. Using the non-orthogonal 
tight binding method, genetic algorithm optimizations 
were carried out for Si$_n$H$_m$ clusters(\emph{n}=1, 2 
and \emph{m}=2-6)\cite{kn:nir1}, \cite{kn:nir2}. Experimental 
studies have been carried out for hydrogenated silicon
clusters using a quadrupole ion trap\cite{kn:hir} where 
Si$_{n}$H$_{x}^{+}$(\emph{n}=2-10 and \emph{x}=0-20) were grown 
from silane gas. From the mass
spectra of these clusters, it was shown how the 
stability of a silicon cluster is  affected by hydrogenation.

Our CPMD calculations show that hydrogen
does not cause any drastic change in  the geometry of the host silicon 
cluster although there is some distortion to the 
structure. To see clearly how the addition of a hydrogen atom 
affects the structure, stability and electronic properties of 
the host silicon cluster, we have also done a number of 
calculations for  host  silicon clusters. We have discussed 
two kinds of stabilities, one is geometrical 
stability and  the other is  electronic. To examine the geometrical
stability of an Si$_{n}$ cluster\cite{kn:car1}-\cite{kn:men}, we have 
calculated the difference between the total energy of 
 modified Si$_{n}$ geometry which 
is obtained by  removal of hydrogen  from Si$_{n}$H 
cluster and ground state energy of Si$_{n}$ cluster. This energy 
difference gives information about the modification of 
the host silicon geometry  due to the addition of hydrogen. 
To examine the electronic stability of Si$_{n}$H cluster 
we have calculated  the first excited 
electronic level gap for Si$_{n}$H clusters. Comparison 
of the first excited electronic level gaps
of Si$_{n}$ clusters and  Si$_{n}$H clusters shows that hydrogen,
in general,  brings electronic stability to silicon 
clusters.  We have also 
calculated  the lowest energy fragmentation products 
of Si$_{n}$ and Si$_{n}$H clusters. 

The plan of the paper is as follows. In section \textbf{II} 
we give computational details of the present work. In 
section \textbf{III} the ground state geometries are presented 
and discussed in detail. In section \textbf{IV} we 
discuss the stability, cohesive energies, total energy 
differences  between 
clusters, the first excited electronic level gaps and 
the lowest energy fragmentation products of Si$_{n}$H 
and Si$_{n}$ clusters. Finally we summarize our 
results in section \textbf{V}.

\section{COMPUTATIONAL DETAILS}

We have used the Car-Parrinello 
 molecular dynamics(CPMD)\cite{kn:car2}, \cite{kn:car3} 
with simulated annealing optimization technique to find 
the ground state structures of  Si$_{n}$ and Si$_{n}$H 
clusters. The CPMD method combines the density functional 
formalism with the molecular dynamics simulation. This scheme 
allows us to describe  dynamics of  ions under  action 
of forces calculated by  
the Hellman-Feynman theorem. The pseudopotentials for silicon 
and hydrogen  
have been generated using the  Bachelet, Hamann
and Schl$\ddot{u}$ter technique\cite{kn:bac}. The local density 
approximation(LDA) of the density functional theory
has been used with the Ceperley-Alder\cite{kn:cep} exchange-correlation 
energy  functional parameterized by Perdew and Zunger\cite{kn:zun}. 

The wave functions were expanded in a plane wave basis with  12 Rydberg
energy cut-off and \textbf{k}=0 point was   
used for Brillouin zone sampling. During  simulation volume of the 
system was kept constant and to avoid  interaction between the 
clusters a big fcc supercell with side length of 35 a.u. 
was used.  To perform simulated annealing, the system was  
taken to high temperatures (1200K in the steps of 300K), equilibiriated 
for a long time (about 16,000 steps)  and 
then slowly cooled down (in the steps of 50K) 
to 300K. Below this temperature the steepest descent optimization 
was found to be  more efficient to obtain the 
ground state geometry. To check the upper limit  
of temperature, some of the clusters were heated up to
1500K and 2000K and it was  found that the resulting ground 
state structures were the same. The 
desired temperature was  achieved by rescaling
 atomic velocities and the  atoms were  moved according to the velocity
Verlet algorithm\cite{kn:tils} with a time step of 5 a.u. The 
fictitious mass of the electron
was  taken  to be  200 a.u. All  calculations were performed 
with more than one initial condition. The initial structures for 
MD calculations were chosen without any pre-assumption about the ground 
state geometry of the cluster. The starting atomic configurations were 
chosen arbitrarily  with a constraint that atoms were neither very far away
from each other nor too close\cite{kn:zhong}, \cite{kn:hohl}. As 
mentioned above these clusters were then heated to high temperatures 
and then equilibiriated for a very long time. At this stage we 
find that the geometry of the hot cluster  does not have any resemblance
with the initial structure. At least two such CPMD calculations were 
performed for each cluster. For some clusters we have done  CPMD 
calculation with three(Si$_9$H) and four( Si$_6$H) starting 
atomic configurations and found that the final 
structures are the same. Furthermore,  we performed the CPMD with steepest
 descent optimizations on the NTBMD structures and
found that the resulting  geometries  either converge to  our 
structures or get stuck in some
local minima.  The ground state structures 
of Si$_{n}$ ( 2\protect\( \leq \emph{n}\leq 10\protect \)) 
clusters obtained by us using the CPMD are similar
to that obtained by previous 
calculations\cite{kn:car1}, \cite{kn:zhong}.  Also our result for
bond length of SiH cluster(1.583\AA) is close to the 
earlier CPMD result\cite{kn:bud1} and other 
calculations\cite{kn:scha}. This shows that our 
calculational procedure is able to give correct structures. 

The first excited electronic level gap 
{\large($\epsilon$ $ _{i+1}$-$\epsilon$$_{i}$)}
 of a cluster is calculated by transferring a small charge 
from its ground state configuration to its first excited state, 
and is given by\cite{kn:von} 
\begin{center}
 {\large$\epsilon$ $ _{i+1}$-$\epsilon$$_{i}$ = $\delta$E$/$ $\delta$q}
\end{center}
where \protect\( \delta \protect \)E is the difference between the 
total energy when \protect\( \delta \protect \)q amount
of charge is  transferred to the first excited state and 
total energy of the ground state.

\section{GROUND STATE STRUCTURES}

In this section we discuss  in detail the ground state geometries 
 of Si$_{n}$H clusters obtained by the CPMD\cite{kn:car2} 
with simulated annealing and steepest descent optimizations. By 
comparing our ground state geometry of Si$_{n}$H 
cluster with Si$_{n}$ cluster, we have investigated the effect 
of hydrogen on 
the geometry of the host silicon cluster. Furthermore, we 
have compared our ground state geometries of Si$_{n}$H clusters 
with earlier calculated geometries of
Si$_{n}$F\cite{kn:rei1} 
and Si$_{n}$Na\cite{kn:rei2} clusters.  We have also made a detailed 
comparison of our work with earlier NTBMD\cite{kn:gup} 
work. It agrees with our geometries of 
Si$_{2}$H, Si$_{3}$H, Si$_{4}$H and Si$_{10}$H  but the  
remaining geometries are different from
our geometries, particularly the position and bonding of
 hydrogen. In the NTBMD\cite{kn:gup} 
results, hydrogen was found to be bonded with more than one 
silicon in most of the clusters but in the present case we find
this only for Si$_{2}$H, Si$_{3}$H and Si$_{5}$H clusters. We find 
that our structures
have lower energies than those of  the NTBMD  
structures.  The nature of bonding 
has been investigated  by performing charge density 
calculations. In the following  we discuss our results for each cluster. 
 
\begin{description}
\item [(a)Si\( _{2} \)H]~
\end{description}

The ground state geometry of Si$_{2}$H cluster is 
shown in Fig. 1(a). Two silicon atoms and  hydrogen form a 
triangular structure. Note that the two silicon atoms are 
bonded to each other not only via Si-Si bond but also via 
Si-H-Si bridge type bond. The lowest  energy structure 
of Si$_{2}$F\cite{kn:rei1} and Si$_{2}$Na\cite{kn:rei2} are 
similar to the structure of Si$_{2}$H. In Si$_{2}$H cluster, Si-Si bond 
length is 2.131\AA\ and  
hydrogen is equidistant from both silicon 
atoms with the bond
length of 1.724\AA\ which  is  larger
than its  value of 1.583\AA\ in Si-H dimer.  The  
Si-Si bond length in Si$_{2}$H cluster is smaller than  
the Si$_{2}$ dimer bond length of 2.184\AA. This implies that  
hydrogen pulls both  silicon atoms  closer and increases 
bonding  between them.  Thus, the additional bonding 
between silicon atoms is due to the Si-H-Si bridge bond 
which is attractive in nature. Such bridge type bonds are 
thought to be present in a:Si-H and play an important 
role in explaining the Staebler-Wronski 
effect\cite{kn:pra1},\cite{kn:fri},\cite{kn:gal},
\cite{kn:stut}. 

It is interesting to note that  hydrogen is bonded
with both  silicon atoms although its  
valance is one. This is  seen clearly from  the valance charge 
density plotted in  Fig. 2,  which shows  existence of 
 bonds between hydrogen and the two silicon atoms.  Such 
overcoordination of hydrogen has also been observed  recently in 
SiC system\cite{kn:gali}. 
We also see from this figure that 
the electrons are more  localized near the 
hydrogen atom. This is expected since hydrogen is 
more electronegative than silicon\cite{kn:paul}. Thus there is a small 
charge transfer from 
silicon atoms to hydrogen\cite{kn:charge}. As a result, Si-H 
bond is neither 
purely covalent nor ionic but is polar covalent\cite{kn:beis}.

\begin{description}
\item [(b)Si\( _{3} \)H]~
\end{description}

The ground state geometry of Si$_{3}$H is shown in  Fig. 1(b). This 
is a planar structure with two fold symmetry and  has some 
resemblance with  Si$_{4}$
cluster\cite{kn:zhong}.  Hydrogen in this cluster 
is bonded with two silicon atoms (1 and 2) and is equidistant from 
both atoms with  bond lengths of 1.715\AA. Silicon atom No.3 is also 
equidistant from silicon atoms 1 and 2. The bond 
length between  silicon 
atoms  1 and 2 is 2.377\AA\ which is more than the Si-Si bond length in 
Si$_{2}$H. This indicates that bonding between 1 and 2 silicon 
atoms is weaker than the Si-Si  bond in Si$_{2}$H because 
of the presence of another silicon atom. Comparing this with 
Si$_{2}$H structure, we note that 
the additional silicon takes diagonal position  opposite 
to hydrogen.  We  see that although hydrogen does not modify 
the  Si$_{3}$ geometry much, it does modify  the 
bond lengths. Particularly, the bond length 
between silicon atoms 1 and 2 in Si$_{3}$H is smaller than
bond length of 2.613\AA\ in Si$_{3}$ cluster. Thus as 
in Si$_{2}$H, hydrogen pulls silicon atoms 1 and 2 closer, which 
can be attributed to the Si-H-Si 
bridge bond. We note that the  
lowest  energy structure of Si$_{3}$Na\cite{kn:rei2} 
is  similar to the present structure, but for Si$_{3}$F cluster
the ground state geometry is  different\cite{kn:rei1}.

\begin{description}
\item [(c)Si\( _{4} \)H]~
\end{description}

The ground state geometry of the cluster is shown 
in  Fig. 1(c). Four silicon 
atoms form a flat rhombus and  the hydrogen atom is 
above the plane and bonded with
one of the silicon atoms. The
same structure was shown as the lowest  energy geometry of Si$_{4}$F
cluster\cite{kn:rei1}. The lowest energy structure of
 Si$_{4}$Na\cite{kn:rei2} cluster is  similar to the  present 
structure but differs in the coordination of Na atom. Comparison of 
this structure 
with Si$_{4}$ structure\cite{kn:zhong} shows that the  
addition of hydrogen does not bring much 
change to Si$_{4}$ structure. Comparison with Si$_{5}$ 
cluster\cite{kn:zhong} shows that this structure does not 
have any resemblance with Si$_{5}$ cluster. Based 
on the idea of  
local softness and hardness, Galv$\acute{a}$n \emph{et al}\cite{kn:joa}
predicted the sites preferred by hard and soft species 
in Si$_{4}$ cluster. We find that hydrogen goes 
to the position according
to their prediction.  

\begin{description}
\item [(d)Si\( _{5} \)H]~
\end{description}

Two lowest energy structures of Si$_{5}$H are very close in energy and 
differ  only by 0.06 eV. The geometry  which is higher in energy is 
shown in  Fig. 1(d). Three of 
the silicon atoms, numbered  1, 4 and 5  form a triangular plane and 
silicon atom No. 2 is above and No. 3  is below  the plane.  Hydrogen 
takes the apex position in the structure and is bonded with  only 
one silicon atom. The geometry of the silicon atoms is same 
as in Si$_{5}$\cite{kn:zhong} cluster. The lowest 
energy structure
of Si$_5$H cluster is shown in  Fig. 1(e). Geometry of this 
cluster is similar to Si$_6$ cluster\cite{kn:zhong} except that 
one of the silicon atoms is replaced by hydrogen. Note that  hydrogen 
in this cluster is attached to two silicon atoms, which is also 
the case in  
Si$_{2}$H and  Si$_{3}$H clusters. Geometrically   hydrogen 
 plays the role of silicon in these three clusters, \emph{i.e.} 
the geometry of Si$_{2}$H is similar to  Si$_{3}$,  Si$_{3}$H
is similar to  Si$_{4}$  and
 Si$_{5}$H is similar to  Si$_{6}$.  We note that a two coordinated 
silicon atom exists in Si$_{3}$, Si$_{4}$ and  Si$_{6}$ clusters and 
hydrogen replaces this silicon atom to form Si$_{2}$H, Si$_{3}$H 
and  Si$_{5}$H respectively. We speculate that this may be a general 
feature of Si$_{n}$H clusters, \emph{i.e.}, if a two coordinated silicon
atoms exists in a Si$_{n+1}$ cluster, hydrogen will replace the two
coordinated silicon atom to form Si$_{n}$H cluster, which will have 
the same geometry as Si$_{n+1}$ cluster. This seems to imply that 
hydrogen will form Si-H-Si bridge type bond between two nearby silicon 
atoms which are doubly coordinated. In the context 
of \emph{a}-Si:H this would 
imply that hydrogen will form Si-H-Si bond between two nearby silicon
atoms having two dangling bonds. Comparison with 
Si$_{5}$F geometry shows that one of the low energy 
structures\cite{kn:rei1} is similar to Si$_{5}$H shown 
in Fig. 1(d). The low energy geometries  for 
 Si$_5$Na\cite{kn:rei2} cluster are similar to 
our low energy geometries for Si$_{5}$H and  the lowest energy 
structure is also the same. 

\begin{description}
\item [(e)Si\( _{6} \)H]~
\end{description}

The ground state geometry of Si$_{6}$H is shown in Fig. 1(f). In this 
structure, four
silicon atoms numbered 1, 2, 3 and 5 form a distorted plane and  
the remaining two atoms numbered 4 and 6 are above the plane. Hydrogen 
atom is bonded to silicon No. 4 from outside the cluster. Comparing 
this with the ground state 
geometry of Si$_{6}$\cite{kn:zhong} cluster, we note 
that although  the  plane formed by four silicon atoms is same   
as in Si$_{6}$ cluster,  the other two silicon  positions 
are different \emph{i.e.} in Si$_{6}$\cite{kn:zhong} 
cluster one silicon atom is above and another is below the Si$_{4}$ 
plane, but in Si$_{6}$H two silicon atoms are above 
the plane. Comparing with the 
results of Si$_{6}$H$_{x}$ clusters\cite{kn:tak}, our 
geometry of Si$_{6}$H
falls in the class of  tetrahedral bonding network. We find that 
this is the only cluster among the clusters considered 
here where the geometry of the silicon  atoms  
differs from the ground state geometry of the 
 host silicon cluster\cite{kn:zhong}. This shows that  
 hydrogen can cause a  transition from  one geometry
to another geometry. One of the low energy geometries 
of Si$_6$F\cite{kn:rei1} cluster is same as the present 
structure. But in the case of 
Si$_6$Na\cite{kn:rei2} one of the low energy structures  has  
similar geometry but it differs from our structure 
by  coordination of the  Na atom.  

\begin{description}
\item [(f)Si\( _{7} \)H]~
\end{description}

The ground state geometry of the structure is shown in 
 Fig. 1(g). Silicon atoms in this structure form a closed and 
compact unit and the cluster has pentagonal symmetry. Five silicon 
atoms numbered  1, 7, 6, 4 and 3 
 make a pentagonal plane and  one  silicon atom is above and another is 
below the plane. Silicon atoms which are not in 
the pentagonal plane are  
bonded to  all the atoms in the pentagonal plane. Hydrogen takes
the apex position in the structure and is bonded with one silicon 
which is out of the
pentagonal plane. It is interesting to note that instead of bonding 
with four coordinated silicon, hydrogen
is bonded with five coordinated silicon. In Si$_{7}$H,
Si$_{8}$H, Si$_{9}$H and Si$_{10}$H
we found the same trend of hydrogen preferring  
to bond with an over-coordinated
silicon atom. This is surprising since  one would have 
expected   it to 
bond with less coordinated
silicon. This may be attributed to slightly higher electronegativity of
H compared to Si and as a result, H  prefers to bond with silicon atom 
having more number of electrons. This is consistent with the  
earlier calculation 
on structural evolution of Si$_{6}$H$_{x}$
clusters\cite{kn:tak}, where it was found that 
hydrogen is not necessarily
bonded with a silicon site having dangling bonds, but with a site
where LUMO amplitude is larger.  Comparison of Si$_{7}$H
with Si$_{7}$ \cite{kn:car1}, \cite{kn:zhong} 
cluster shows  that hydrogen hardly changes the geometry of Si$_{7}$
cluster implying that Si$_{7}$ cluster is 
a very stable cluster. The lowest energy geometry 
of Si$_7$F\cite{kn:rei1} and one of the low energy structure 
of Si$_7$Na\cite{kn:rei2} is similar to our ground state geometry. 

\begin{description}
\item [(g)Si\( _{8} \)H]~
\end{description}

Figs. 1(h) shows  the  ground state 
geometry of Si$_{8}$H. This
structure is also made of a compact, closed unit of  silicon 
atoms with  hydrogen sticking to the structure
from outside. We see that  Si$_{8}$H shows some 
similarity with Si$_{7}$H
cluster \emph{i.e.} the same pentagonal plane formed by silicon 
atoms numbered 4, 3, 1, 7 and 2 exists in Si$_{7}$H and 
 one  silicon is  above and another is below the 
plane(6 and  5) as in
Si$_{7}$H. Silicon atom 8 is attached to the
triangular plane of the Si$_{7}$H structure in such a way that
it is away from  hydrogen atom. As in Si$_{7}$H  hydrogen atom 
is attached to the silicon atom 
which is bonded with five silicon atoms. Fig. 1(h) shows that  
Si$_{8}$H has two distorted
Si$_{4}$ planes; silicon atoms numbered 8, 4, 3 and 5 form 
one distorted 
Si$_{4}$ plane and 2, 6, 1 and 7 form another. These planes are 
not parallel but rotated with respect to each other
in such a way as to have more than one bond 
for each silicon atom with the atoms in 
the other plane.  Comparing our ground state geometry of Si$_{8}$H with
Si$_{8}$\cite{kn:car1} geometry, we see that silicon atoms have 
similar geometry except that the  Si$_{4}$
planes are distorted in  Si$_{8}$H.

\begin{description}
\item [(h)Si\( _{9} \)H]~
\end{description}

The ground state geometry of the cluster is shown in  Fig. 1(i). This 
structure is also compact and closed by silicon atoms. The structure 
consists of two Si$_{4}$ planes formed by atoms  1, 3, 4, 9 and 
5, 8, 2, 7 and  silicon No.6  forms a cap. Hydrogen is connected to 
silicon atom No. 9 which is coordinated with five 
silicon atoms. Comparison 
with Si$_{9}$\cite{kn:car1} cluster shows that the atoms forming
Si$_{4}$ planes in Si$_{9}$H do not lie in a plane 
in Si$_{9}$. 

\begin{description}
\item [(i)Si\( _{10} \)H]~
\end{description}

The ground state geometry of this cluster is shown in Fig. 1(j).  The 
Si$_{10}$ structural unit in Si$_{10}$H is similar to  Si$_{9}$H 
cluster  and the additional silicon(atom 1) 
makes a side cap to Si$_{9}$H cluster( Fig. 1(j)). Also there are 
two Si$_{4}$ planes rotated with respect to each other as in Si$_{8}$H
and Si$_{9}$H. Silicon atoms  5, 2, 7 and 6  form one plane
and 8, 3, 10 and 4 form other plane. In this cluster 
 hydrogen is connected to the five 
fold coordinated silicon. 
Comparison with Si$_{10}$ \cite{kn:car1} 
cluster shows  that hydrogen hardly changes the geometry of Si$_{10}$
cluster implying that Si$_{10}$ cluster is a very stable cluster. A
general feature of clusters Si$_{6}$H, Si$_{7}$H, Si$_{8}$H, Si$_{9}$H
and Si$_{10}$H is that silicon atoms in the cluster
form a closed compact unit with
hydrogen outside this structural unit.

\section{STABILITY OF Si$_{n}$ AND Si$_{n}$H CLUSTERS}

We find that the total energy of Si$_{n}$ as well as of  Si$_{n}$H clusters 
increases approximately linearly with the cluster size \emph{n}. The 
addition of hydrogen 
to an Si$_{n}$ cluster reduces the energy of  the cluster by
approximately 15 eV. The cohesive energy 
per particle versus  number of silicon atoms is plotted in
 Fig. 3. As seen clearly from  the figure cohesive energy per particle 
increases rapidly  up to Si$_6$H cluster and then it 
increases slowly as a function of size. As noted earlier, from this 
size ( Si$_6$H) onwards  silicon atoms in the
cluster form closed compact unit and some of the silicon atoms 
have  coordination number more than four. 

We take the   first excited electronic
level gap of a cluster as the  difference between  the  
first excited electronic level and the highest occupied level. For  
closed-shell or sub-shell systems  
 this gap will be  the same as HOMO-LUMO gap which   
 is related to the chemical hardness and  electronic stability of 
a system\cite{kn:juv}-\cite{kn:har}. Though 
the first excited 
electronic level gap is not equivalent to HOMO-LUMO gap for Si$_{n}$H
clusters, it can be related to  electronic 
stability.  A bigger value of the first excited electronic level gap 
for a system means that  it is difficult to excite electrons 
from its ground state  and thus
the electronic system can sustain its ground state for 
larger perturbations.  Thus the first excited 
electronic level gap can be taken as a measure of  the
electronic stability of a system. We have shown the 
first excited level gap
as a function of cluster size  in Fig. 4 for Si$_{n}$H and Si$_{n}$
clusters. Also shown in the figure are results 
of Lu et al\cite{kn:zhong}
for Si$_{n}$ clusters which are in good agreement with our results. We 
see that general trend of variation of first excited electronic 
level gap is quite similar for Si$_{n}$H 
and Si$_{n}$ clusters. The figure also shows that the addition 
of hydrogen can cause large changes in the electronic structure of 
Si$_{n}$ cluster.

From Fig. 4  we see that the gap fluctuates with 
size, which indicates that the gap strongly depends on  the size
and geometry of a  cluster. It might be interesting to 
draw parallels with 
short-range potential fluctuations in \emph{a}-Si:H 
system which occur at the length length scale of 3\AA\cite{kn:sca3}. It 
can be argued that an amorphous system can be considered 
as a loosely connected 
network of small clusters and thus our calculation provoids a 
first-principle basis for the potential fluctuations 
\cite{kn:fri1}-\cite{kn:sca3}.  Furthermore, we see that the first 
excited electronic level gap for Si$_{n}$H is, on an average, larger than 
that of Si$_{n}$ cluster. This is consistent with the observation that 
the band gap of \emph{a}:Si increases on 
hydrogenation\cite{kn:cody}.  Further, Fig. 4 shows that 
Si$_{2}$H, Si$_{3}$H, Si$_{5}$H, 
Si$_{7}$H, Si$_{9}$H and Si$_{10}$H
clusters are  electronically more stable compared to 
 Si$_{4}$H, Si$_{6}$H and  Si$_{8}$H clusters.  Also we see 
that Si$_{2}$, Si$_{5}$, Si$_{6}$, Si$_{7}$ and Si$_{10}$ clusters 
are   electronically more
stable than  other silicon clusters(Si$_{3}$, Si$_{4}$, Si$_{8}$ 
and  Si$_{9}$ clusters), since they have larger gaps. 

 To examine geometrical stability we have calculated  
the difference 
between the total energy  of the modified Si$_{n}$ 
cluster, which has the 
same positions of 
silicon atoms as in the Si$_{n}$H cluster, and the 
ground state geometry 
of Si$_{n}$ cluster.  This energy difference 
is a measure of how much a  silicon 
cluster distorts  from its ground state geometry 
due to the addition of hydrogen atom. Lower value of this difference 
for a Si$_{n}$ cluster  
means that the cluster is geometrically  stable. This total
energy difference as a function of cluster size \emph{n} is shown 
in  Fig. 5. The figure shows that
Si$_{2}$, Si$_{4}$, Si$_{7}$ and Si$_{10}$ clusters 
are geometrically  more stable 
than Si$_{3}$, Si$_{5}$, Si$_{6}$, Si$_{8}$ and  Si$_{9}$ 
clusters.  On the other
hand Si$_{3}$, Si$_{5}$, Si$_{6}$, Si$_{8}$ and Si$_{9}$ clusters are
stabilized by hydrogen and have a greater tendency to
adsorb hydrogen.  This is consistent with the conclusions 
drawn by comparing Si$_{n}$H and  Si$_{n}$ ground state 
geometries in section \textbf{III}, as  Si$_{2}$, Si$_{4}$, Si$_{7}$
and Si$_{10}$ clusters were least distorted  by the addition 
of hydrogen. Thus the above discussion shows that,
 Si$_{2}$, Si$_{7}$  and Si$_{10}$ clusters are the most 
stable clusters from both 
viewpoints of electronic as well as geometrical stability. 

As pointed out earlier in section \textbf{III}, hydrogen is 
attached to  silicon clusters from outside in 
several cases (Si$_{4}$H to Si$_{10}$H). To examine  
this further, we have performed the steepest descent 
optimization on Si$_{6}$H and Si$_{7}$H
clusters with hydrogen surrounded by silicon atoms.  We find  that  
 hydrogen atom always comes out of the silicon cluster independent of 
 the cluster size. This is mainly due to the higher 
electrostatic energy 
of the cluster when hydrogen is inside the cluster.  Thus our result 
implies that hydrogen will tend 
to come out of  crystalline silicon and would like 
to stay on the surface. This is consistent with the 
experimental observation 
in which  hydrogen is used to produce
 homogeneous  silicon surface by terminating  surface 
silicon dangling bonds to reduce the
surface reconstruction\cite{kn:gal}. 

To investigate  fragmentation products of Si$_{n}$ 
and Si$_{n}$H clusters, 
we have calculated  the difference between the total energy of a 
cluster which undergoes fragmentation and that of its possible product 
clusters. The most probable pathway for fragmentation  of  a particular 
cluster is  the one which has the smallest 
total energy difference\cite{kn:rhag}.  Since the clusters are small 
in size, we are assuming 
that the fragmentation results  in only  two
product clusters. Our calculations are only for neutral 
fragmentation of Si$_{n}$ and Si$_{n}$H clusters.  Our lowest energy 
fragmentation products of Si$_{n}$ 
clusters agree very well with all primary fragmentation 
products calculated  
by Shvartsburg \emph{et al}\cite{kn:shv}. In Table. 1 we have given 
the lowest energy 
fragmentation products of Si$_{n}$H clusters with the corresponding 
dissociation energies. We see from the table that 
the lowest energy fragmentation products have hydrogen atom as one 
of the products for all Si$_{n}$H clusters except for Si$_{8}$H  
cluster. This shows that it is easy to remove hydrogen  from 
Si$_{n}$H clusters. 

\section{CONCLUSIONS}

We have presented detailed  results for the ground state structures and 
electronic properties of Si$_{n}$H clusters using the Car-Parrinello 
molecular dynamics simulations. We find that hydrogen can form a
 bridge like Si-H-Si bond connecting two silicon atoms. Such bridge
like bonds are thought to be present 
in \emph{a}-Si:H\cite{kn:pra1}, \cite{kn:fri}, 
\cite{kn:stut}. However, among 
the clusters considered here hydrogen
forms  a bridge like bond only in Si$_{2}$H, Si$_{3}$H 
and Si$_{5}$H clusters; in others it is bonded with
only one silicon atom  and attached to the cluster from 
outside.  Charge density calculations show that 
the  Si-H bond in all clusters is polar covalent. In  
clusters from Si$_{7}$H to Si$_{10}$H, silicon atoms form a 
compact unit and hydrogen attaches to a silicon atom 
which is over-coordinated. Though hydrogen has 
small effect on the  geometry
of the host silicon cluster, it changes bond lengths and 
tries to distort the silicon cluster. This is similar to the 
behavior of hydrogen in \emph{a}-Si:H 
where it has been found that hydrogen creates local distortions as
it moves\cite{kn:pra1}, \cite{kn:fri}. We find that hydrogen 
has a tendency to come out of compact silicon clusters and 
prefers to stay out of the cluster. This is 
consistent with the behavior of hydrogen on silicon 
surfaces\cite{kn:gal}. 

The first excitation electronic level gap of 
the Si$_{n}$H clusters fluctuates as a function of size 
and this may provide
a first principles basis for the short range potential fluctuations in 
\emph{a}-Si:H\cite{kn:fri1}-\cite{kn:sca3}. Our calculations show that 
the addition of hydrogen can cause large changes in the electronic 
structure of host Si$_{n}$ cluster. Furthermore, it shows that 
Si$_{2}$H, Si$_{3}$H, Si$_{5}$H, Si$_{7}$H, Si$_{9}$H 
and Si$_{10}$H clusters
are electronically more stable than Si$_{4}$H, Si$_{6}$H 
and  Si$_{8}$H clusters. We find that Si$_{2}$, Si$_{4}$, Si$_{7}$ 
and Si$_{10}$ clusters are geometrically  more stable 
than Si$_{3}$, Si$_{5}$, Si$_{6}$, Si$_{8}$ and  Si$_{9}$ 
clusters, while  Si$_{2}$, Si$_{5}$, Si$_{6}$, Si$_{7}$ and  Si$_{10}$ 
clusters are electronically more stable 
than Si$_{3}$, Si$_{4}$,  Si$_{8}$ 
and Si$_{9}$ clusters.  We have calculated the lowest 
energy fragmentation products of Si$_{n}$ and Si$_{n}$H 
clusters. Our results
for fragmentation products of Si$_{n}$ clusters agree very 
well with the earlier
predictions. The lowest energy fragmentation products 
of Si$_{n}$H clusters
show that it is easy to remove hydrogen  from 
silicon clusters. Comparison 
of  Si$_{2}$H to Si$_{7}$H cluster 
with the corresponding F and Na substituted clusters 
shows that almost all 
have similar low energy geometries implying that the geometrical 
effect of H, F and Na on Si$_{n}$clusters are similar.

\begin{center}
\textbf{Acknowledgment}
\end{center}

 It is a pleasure to thank Drs. S. C. Agarwal, V. A. Singh, 
M. K. Harbola, Y. N. Mohapatra, A. Bansil and Roy Benedek 
for helpful discussions 
and comments. This work was supported by the Department of Science 
and Technology, New Delhi via project No. SP/S2/M-51/96.

\newpage
Table. 1. Fragmentation pathways of a neutral Si\( _{n} \)H cluster 
into products
Si\( _{m} \) and Si\( _{n-m} \)H cluster. 

\vspace{0.3cm}
{\centering \begin{tabular}{|c||c|c||c|}
\hline 
reactant&
product&
product&
dissociation\\
clusters Si\( _{n} \)H&
cluster Si\( _{m} \)&
cluster Si\( _{n-m} \)H&
energy in eV\\
\hline
SiH&
Si&
H&
4.60\\
\hline
Si\( _{2} \)H&
Si\( _{2} \)&
H&
4.73\\
\hline
Si\( _{3} \)H&
Si\( _{3} \)&
H&
3.97\\
\hline
Si\( _{4} \)H&
Si\( _{4} \)&
H&
3.29\\
\hline
Si\( _{5} \)H&
Si\( _{5} \)&
H&
3.38\\
\hline
Si\( _{6} \)H&
Si\( _{6} \)&
H&
3.17\\
\hline
Si\( _{7} \)H&
Si\( _{7} \)&
H&
2.57\\
\hline
Si\( _{8} \)H&
Si\( _{7} \)&
SiH&
3.25\\
\hline
Si\( _{9} \)H&
Si\( _{9} \)&
H&
3.04\\
\hline
Si\( _{10} \)H&
Si\( _{10} \)&
H&
2.96\\
\hline
\end{tabular}\par}
\vspace{0.3cm}

\vspace{0.3cm}
{\par}
\vspace{0.3cm}

\newpage
\begin{center}
\Large\bf{FIGURE CAPTIONS}
\end{center}
\vskip .2in

{\bf Fig. 1}. Ground state geometry of {\bf(a)} Si$_2$H, 
{\bf(b)} Si$_3$H, {\bf(c)} Si$_4$H. {\bf(d)} Higher 
energy geometry of Si$_5$H. 
{\bf(e)} Ground state  geometry of Si$_5$H, {\bf(f)}  
Si$_6$H, {\bf(g)} Si$_7$H, {\bf(h)} Si$_8$H, 
{\bf(i)} Si$_9$H and  {\bf(j)} Si$_{10}$H cluster. Silicon atoms are
numbered and hydrogen atom is shown by a small dark circle. A thick line
between two atoms indicates a bond between the atoms.  

{\bf Fig. 2}. Valence charge density of Si$_2$H cluster in 
arbitrary units plotted in the plane of the cluster. Constant 
charge density contours are also shown. Approximate  
positions of Si and H atoms are indicated by arrows.

{\bf Fig. 3}. Cohesive energy per atom of Si$_{n}$H  cluster versus 
number of silicon atoms.

{\bf Fig. 4}. First excited state electronic level gap of  Si$_{n}$H 
 and Si$_{n}$ cluster versus number of silicon atoms in 
the cluster. The circles and squares  correspond 
to Si$_{n}$H and Si$_{n}$ clusters
respectively. The triangles  represent the results of 
Lu \emph{et al}\cite{kn:zhong}
for Si$_{n}$ cluster.

{\bf Fig. 5}. Difference dE,  between total energy of  
 the modified geometry of Si$_{n}$ and the ground state 
energy of Si$_{n}$ 
cluster versus number of silicon atoms in the clusters. 


\begin{thebibliography}{99}
\bibitem{kn:foot} The ratio of computational effort involved in CPMD
and NTBMD calculation is about 2 orders of magnitude.
\bibitem{kn:paul}On Pauling scale, electronegativity of hydrogen and 
silicon is 2.2 and 1.9 respectively. See, for example, 
Ref. \cite{kn:beis}
\bibitem{kn:charge}Charge transfer is difficult to calculate 
as it is not a well defined quantity. To estimate  
roughly the charge transfer from silicon to 
hydrogen in Si$_{2}$H first we  calculated amount 
 of charge in  cube  of length  2.0 a.u with  hydrogen  at the center 
of the cube. Similar calculation was repeated for a single 
free H atom. The  difference between the amount of charge in 
two cases shows approximately 0.25 electron transfer from Si$_{2}$ to 
hydrogen atom.
\end{thebibliography}
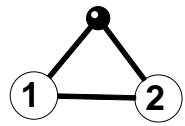
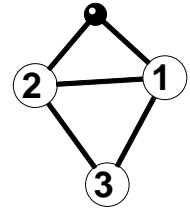
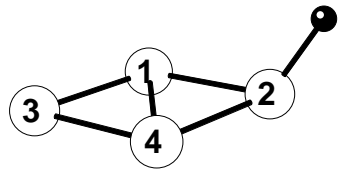
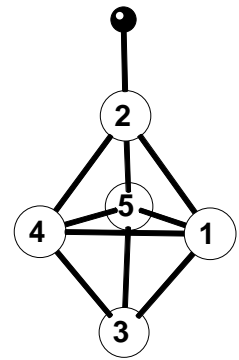
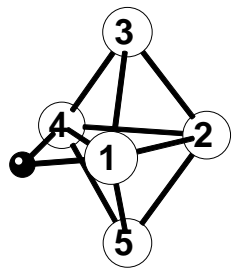
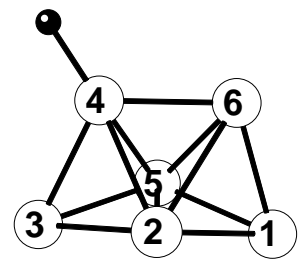
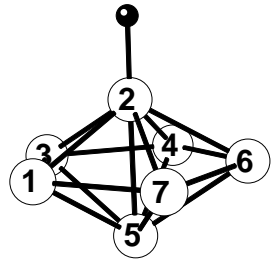
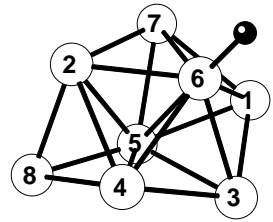
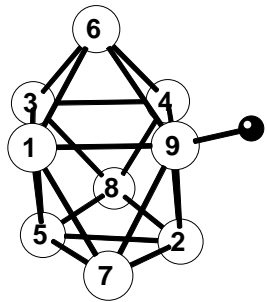
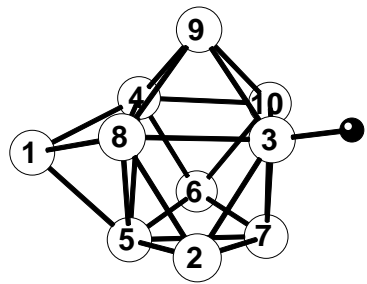
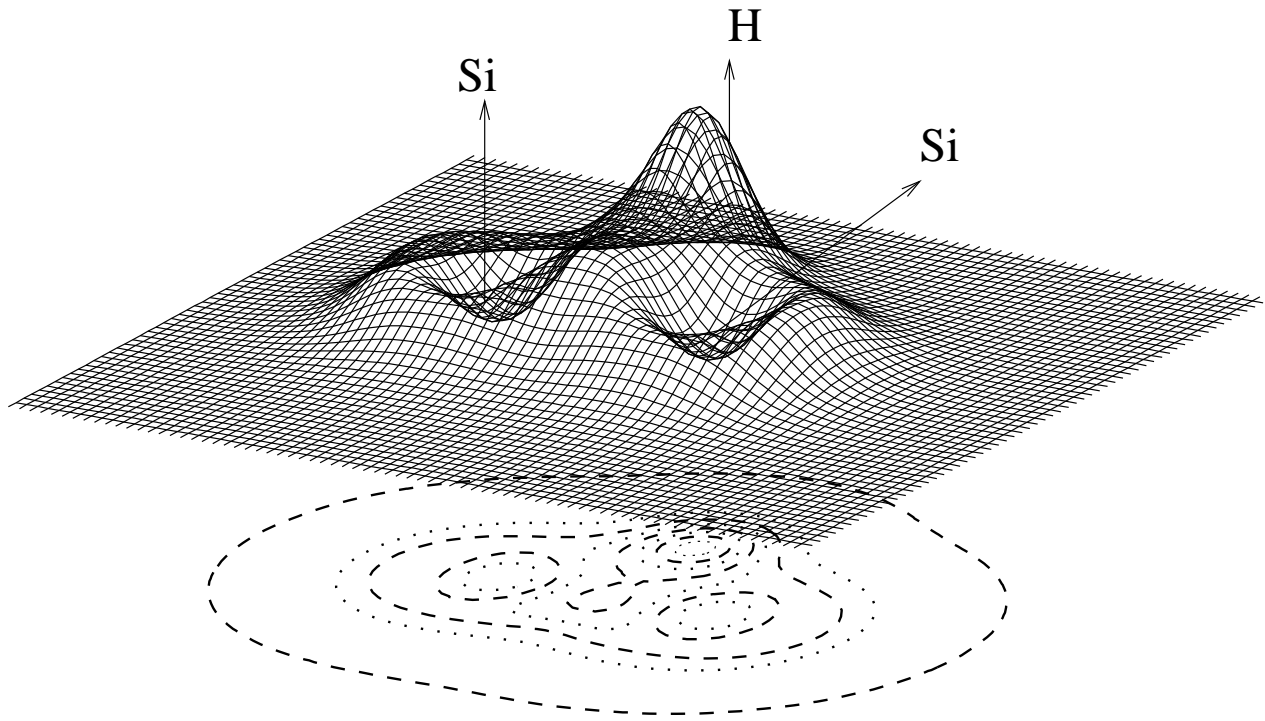
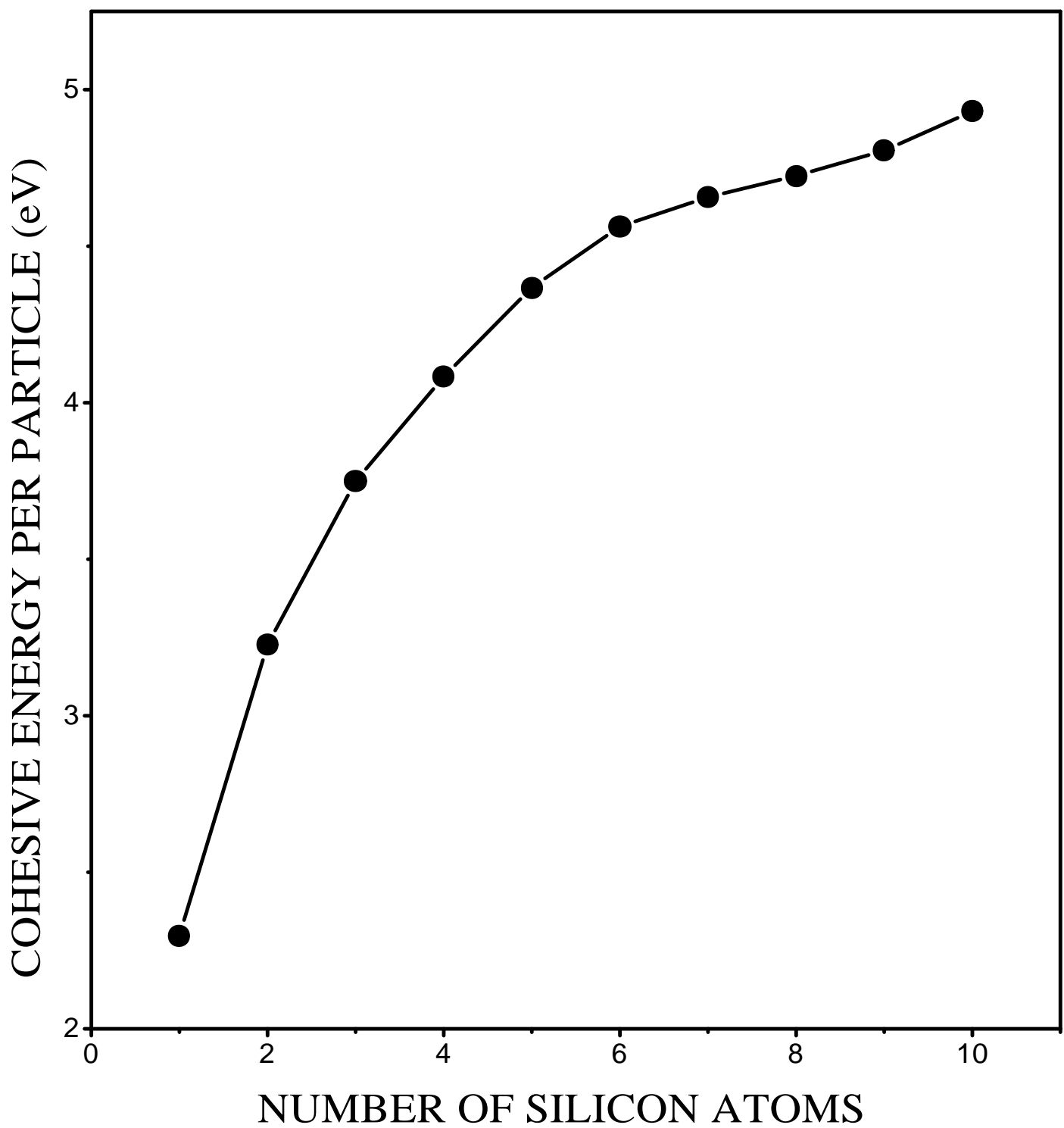
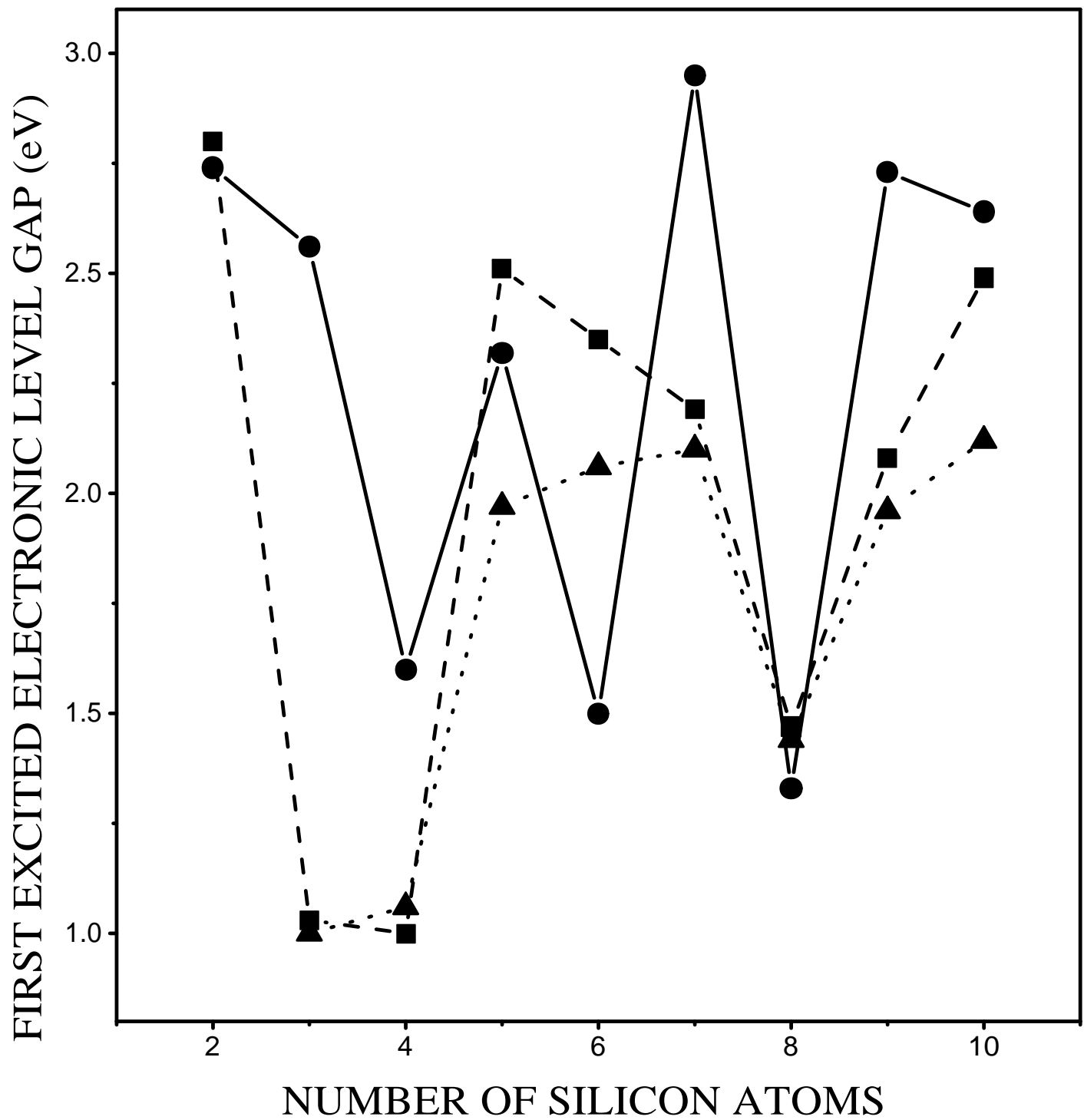
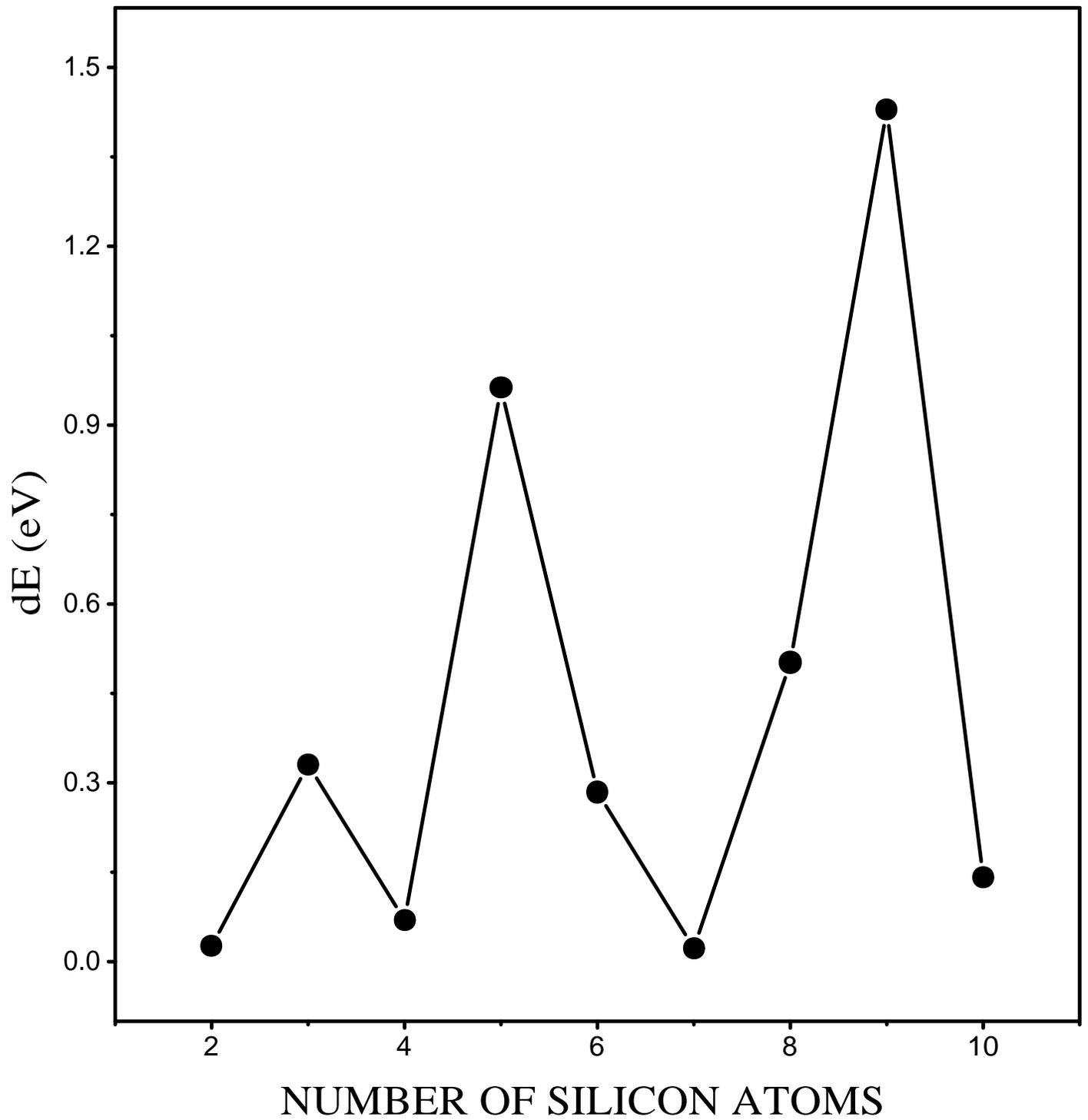
\end{document}